# Variational Resummation of Divergent Series with known Large-Order Behavior

H. Kleinert*

*Institut für Theoretische Physik,*
*Freie Universität Berlin*
*Arnimallee 14, D - 14195 Berlin*

## Abstract

Recently-developed variational perturbation expansions converge exponentially fast for positive coupling constants. They do not, however, possess the correct left-hand cut in the complex coupling constant plane, implying a wrong large-order behavior of their Taylor expansion coefficients. We correct this deficiency and present a method of resumming divergent series with their proper large-order behavior. For a given set of expansion coefficients, knowledge of the large-order behavior considerably improves the quality of the approximation.

Typeset using REVTEX

---

*email: kleinert@einstein.physik.fu-berlin.de; URL: http://www.physik.fu-berlin.de/~kleinert

1) Recently, the Feynman-Kleinert variational approximation to path integrals [1] has been extended to a systematic variational perturbation expansion [2,3] which converges uniformly and fast (for the anharmonic oscillator like $e^{-\text{const}\times N^{1/3}}$ in the order $N$ of the approximation) [4,5]. Due to the uniformity of the convergence, this extension has given rise to an efficient method for extracting strong-coupling expansions from a weak-coupling expansions [6–8]. This has led to a new resummation procedure which makes use of a possible independent knowledge of expansion coefficients for weak and strong couplings [9], and should prove useful for extracting critical properties from lattice models of statistical mechanics, where both expansions are known to high order (see for example the testbooks [10]).

In the theory of critical exponents, a different resummation procedure is required. There one must evaluate divergent series of which one typically knows only a few expansion coefficients plus the large-order behavior, the latter from semiclassical tunneling theory [3,11]. We present a method of combining the exponentially fast convergence of variational perturbation theory with the information on the large-order behavior, thus laying the grounds for an efficient resummation of perturbative expressions in the quantum field theory of critical phenomena.

The method is completely general and holds for any physical system whose quantities, for instance energy eigenvalues $E$, possess (possibly divergent) power series expansions in some dimensionless coupling constant $g' = g/\omega^q$ ($\omega$ being some energy parameter) of the type $\omega^p \sum E_n (g/\omega^q)^n$, whose coefficents behave for large orders like

$$E_k = \gamma p^{\beta+1} k^\beta (-a)^k (pk)! \left[ 1 + \frac{\gamma_1}{k} + \frac{\gamma_2}{k^2} + \ldots \right]. \tag{1}$$

Such a behavior arises from a cut in the complex coupling constant plane, across which $E(g)$ has a discontinuity

$$\text{disc } E(-|g|) \equiv E(-|g| - i\eta) - E(-|g| + i\eta) = 2i\text{Im}E(-|g| - i\eta) \tag{2}$$

$$= (a|g'|/\omega^q)^{-(\beta+1)/p} e^{-1/(a|g'|)^p} [1 + c_1 (a|g'|)^{1/p} + c_2 (a|g'|)^{2/p} + \ldots]. \tag{3}$$

A typical example is the ground state energy $E(g)$ of the anharmonic oscillator which has $p = 1, q = 3$ and a large-order behavior

$$E_k = -\frac{\omega}{\pi}\sqrt{\frac{6}{\pi}}(-3/\omega^3)^k\Gamma(k+1/2)[1-95/72k+\ldots],\tag{4}$$

with a discontinuity [12]

$$2i\mathrm{Im}E(-|g|-i\eta) = -2i\omega\sqrt{\frac{6}{\pi}}\sqrt{\frac{\omega^3}{3|g|}}e^{-\omega^3/3|g|}[1-(95/72)(3|g|)+\ldots].\tag{5}$$

In this note, we shall develop the method for this particular example and illustrate how the information on the large-order behavior accelerates the convergence of variational perturbation expansions.

2) Following the procedure explained in [3], we take the weak coupling expansion of order $N$,

$$E_N = \omega^p \sum_{n=0}^{N} E_n \left(\frac{g}{\omega^q}\right)^n,\tag{6}$$

and replace $\omega$ by the identical expression

$$\omega \to \sqrt{\Omega^2 + \omega^2 - \Omega^2}\tag{7}$$

and reexpand $E_N^w$ in powers of $\lambda$ treating $\omega^2 - \Omega^2$ as a quantity of order $g$. The reexpanded series is truncating after the order $n > N$.

The resulting expansion has the form

$$W_N(g,\Omega) = \Omega^p \sum_{n=1}^{N} a_n f_n(\Omega) \left(\frac{g}{\Omega^q}\right)^n\tag{8}$$

where

$$f_n(\Omega) = \sum_{j=0}^{N-n} \binom{(p-qn)/2}{j}(-)^j\left(1-\frac{\omega^2}{\Omega^2}\right)^j.\tag{9}$$

Forming the first and second derivatives of $W_N(g,\Omega)$ with respect to $\Omega$ we find the positions of the extrema and the turning points. The smallest among these is denoted by $\Omega_N$. The resulting $W_N(g) \equiv W_N(g,\Omega_N)$ constitutes the desired approximation to the energy.

3) The perturbation expansion of the anharmonic oscillator looks like (6) with $E_n = 1/2, 3/4, -21/8, 333/16, -30885/128, 916731/256\ldots$. The lowest-order approximation to the energy reads

$$W_1(g,\Omega) = \left(\frac{\Omega}{2} + \frac{1}{2\Omega}\right) E_0 + E_1 \frac{g}{\Omega^2}. \tag{10}$$

Extremizing this yields

$$\Omega_1 = \begin{cases} \frac{2}{\sqrt{3}}\omega \cosh\left[\frac{1}{3}\mathrm{acosh}(g/g_1)\right] & g > g_1, \\ \frac{2}{\sqrt{3}}\omega \cos\left[\frac{1}{3}\arccos(g/g_1)\right] & g < g_1, \end{cases} \quad \text{for} \tag{11}$$

with $g_1 \equiv 2\omega^3 E_0/3\sqrt{3}E_1$. The result is shown in Fig. 1, where the approximation is seen to have a maximal error of 2% for large couplings.

For larger negative couplings $g < 0$, the imaginary part of the energy is reproduced with the same type of error, as shown in Fig. 2. At small negative $g$, however, the approximation $W_1(g)$ has an important qualitative deficiency: It does not reproduce the imaginary part (5) in the interval $g \in (-g_1, 0)$ where $g_1 \approx 0.2566$. known from semiclassical tunneling theory.

By going on to the approximation $W_3(g)$, the energy at positive $g$ is found correctly within 0.2% (see Figs. 1 and 3). The same is true for the imaginary part at larger negative $g$ (see Figs. 2, 4, and 5). Again, the imaginary part at small negative $g$ is being missed, although the interval is now smaller: $g \in (-g_3, 0)$ with $g \approx 0.16$

4) Let us now include the knowledge of the imaginary part at small $g < 0$ in Eq. (5), derived via tunneling theory, into the variational perturbation expansion. First we consider the approximation $W_1(g)$ where the missing interval has $g_1 = 4/9Sqrt3 \approx 0.2566$. The dispersion relation for $E(g)$ reads (with one subtraction to ensure convergence)

$$E(g) = \frac{\omega}{2} + 2\omega g \int_0^\infty \frac{d\lambda}{2\pi} \frac{1}{\lambda(\lambda+g)} \sqrt{\frac{6}{\pi}} \sqrt{\frac{\omega^3}{3\lambda}} e^{-\omega^3/3\lambda} \varepsilon^{\mathrm{i}}(g) \tag{12}$$

where $\varepsilon^{\mathrm{i}}(g)$ starts out like $1 - (95/72)(3|g|) + \ldots$ and looks as shown in Fig. 2. By expanding $1/(\lambda+g)$ in a power series in $g$, we obtain the expansion coefficients as the moment integrals of the imaginary part as a function of $1/g$:

$$E_k = -2\omega \int_0^\infty \frac{d\lambda}{2\pi} \frac{1}{\lambda^{k+1}} \sqrt{\frac{6}{\pi}} \sqrt{\frac{\omega^3}{3\lambda}} e^{-\omega^3/3\lambda} \varepsilon^{\mathrm{i}}(g) \tag{13}$$

We now assume only the knowledge of the leading semiclassical imaginary part (5). This is used to approximate the imaginary part in the entire regime where $W_1(g)$ is real for negative $g$, i.e., in the interval $g \in (-g_1, 0)$. There it contributes to the expansion coefficients

$$\Delta_1 E_k = -2\omega \int_0^{g_1} \frac{d\lambda}{2\pi} \frac{1}{\lambda^{k+1}} \sqrt{\frac{6}{\pi}} \sqrt{\frac{\omega^3}{3\lambda}} e^{-\omega^3/3\lambda} \varepsilon^{\mathrm{i}}(g) \tag{14}$$

These numbers are subtracted from the full expansion coefficients $E_k$ forming $E'_k$. For the new coefficients $E'_k$, the imaginary part of $W_1(g)$ starts out at another value of $g_1 \equiv 2\omega^3 E'_0 / 3\sqrt{3} E'_1$, for which we calculate from (13) new coefficients $E'_k$, and so on. This procedure converges at $g_1 \approx 0.166$. For the associated coefficients $E'_k \approx 0.50117, 0.72905, -2.24059, 13.54295, -98.64571, \ldots$ we now evaluate the variational perturbation expansion $W'_1(g)$. To this we add the energy associated with the dispersion integral (14):

$$\Delta_1 E(g) = 2\omega g \int_{g_1}^\infty \frac{d\lambda}{2\pi} \frac{1}{\lambda + g} \sqrt{\frac{6}{\pi}} \sqrt{\frac{\omega^3}{3\lambda}} e^{-\omega^3/3\lambda} \varepsilon^{\mathrm{i}}(g). \tag{15}$$

No subtraction is necessary. For positive $g$, the new approximation $\bar{W}'_1(g) \equiv'_1(g) + \Delta_1 E(g)$ is shown in Fig. 1. It is seen to be better roughly by about 30% than the previous approximation $W_1(g)$.

The important qualitative advantage of the approximation $\bar{E}_1(g)$ is seen in Fig. 2. The imaginary part starts now at $g = 0$, remains constant until $g = -g_1$, and is approximated by Im $W'_1(g)$ for $g < -g_1$.

We now go on to the approximation $W_3(g)$ which for $g > 0$ is accurate to 0.05% as shown in Fig. 3. The situation for the imaginary part at larger negative $g$ is only slightly worse (see Figs. 4 and 5). The imaginary part is now missing in the interval $g \in (-g_1, 0)$ with $g_1 \approx -0.16$. For this $g_1$, we evaluate the dispersion relation (13) and find new expansion coefficients $E'_k$. These render a new value of $g_1$, etc., until the method converges at $g_1 \approx 0.166$. The resulting new expansion coefficients are

$E'_k \approx 0.5000477, 0.74871, -2.58993, 19.84402, -214.12062, \ldots$ . To the associated $W'_3(g)$ we add the contribution from the dispersion integral (14), and obtain the final result $\bar{W}'_3(g) \equiv '_3(g) + \Delta_3 E(g)$. In Fig. 3 we see that for $g > 0$, the new approximation is better than the previous one $W_3(g)$ by roughly a factor 5. To see the convergence of the iteration in $g_1$, we also plot $W'_3(g)$ for the initial $g_1 \approx 0.16$.

The important qualitative advantage of the new approximation is visible in Figs. 4 and 5. There is now an imaginary part for all negative $g$ which ensures the correct leading large-order behavior of the new approximation (4). The cut in the interval $g \in (-0.166, 0)$ is approximated by leading term in the semiclassical expression (5).

In the special case of the anharmonic oscillator, the convergence could of course be improved by using our knowledge of correction factor for the imaginary part in Eq. (5). However, since such correction factors are generally hard to derive, we have not made use of them.

5) The new resummation method incorporates into the variational perturbation expansion the initial, tunneling section of the left-hand cut, thus accounting for the correct large-order behavior of the expansion coefficients. This should prove useful in calculations of critical exponents of statistical systems described by quantum field theory, where the tip of the cut is known from tunneling theory. It will be interesting to see whether it will be possible to improve upon the results obtainen by Borel-Padé methods.

Certainly, any knowledge of the strong-coupling behavior can be used to obtain better approximations along with the help of the procedure developed in Ref. [9].

Note that because of the scaling relation

$$E(g,\omega) = \omega E(g/\omega^3, 1) \tag{16}$$

first noted by Symanzik, the resummation problem for the non-Borel-summable double-well potential with $\omega^1 = -1$ can be treated in the same way [14].

# REFERENCES

[1] R.P Feynman and H. Kleinert, Phys. Rev. **A34**, 5080 (1986).

[2] H. Kleinert, Phys. Lett. **A173**, 332 (1993).

[3] See Chapters 5 and 17 in the textbook

H. Kleinert, *Path Integrals in Quantum Mechanics, Statistics and Polymer Physics*, World Scientific, Singapore 1995.

[4] Related expansions have been proposed by

R. Seznec and J. Zinn-Justin, J. Math. Phys. **20**, 1398 (1979); T. Barnes and G.I. Ghandour, Phys. Rev. **D22**, 924 (1980); B.S. Shaverdyan and A.G. Usherveridze, Phys. Lett. **B123**, 316 (1983); H. Mitter and K. Yamazaki, J. Phys. A **17**, 1215 (1984);

P.M. Stevenson, Phys. Rev. **D30**, 1712 (1985); **D32**, 1389 (1985); P.M. Stevenson and R. Tarrach, Phys. Lett. **B176**, 436 (1986); A. Okopinska, Phys. Rev. **D35**, 1835 (1987); **D36**, 2415 (1987); W. Namgung, P.M. Stevenson, and J.F. Reed, Z. Phys. **C45**, 47 (1989); U. Ritschel, Phys. Lett. **B227**, 44 (1989); Z. Phys. **C51**, 469 (1991); M.H. Thoma, Z. Phys. **C44**, 343 (1991); I. Stancu and P.M. Stevenson, Phys. Rev. **D42**, 2710 (1991); R. Tarrach, Phys. Lett. **B262**, 294 (1991); H. Haugerud and F. Raunda, Phys. Rev. **D43**, 2736 (1991); A.N. Sissakian, I.L. Solivtosv, and O.Y. Sheychenko, Phys. Lett. **B313**, 367 (1993).

[5] First convergence proofs by

I.R.C. Buckley, A. Duncan, and H.F. Jones, Phys. Rev. **D47**, 2554 (1993); C.M. Bender, A. Duncan, and H.F. Jones, Phys. Rev. **D49**, 4219 (1994); A. Duncan and H.F. Jones, Phys. Rev. **D47**, 2560 (1993); R. Guida, K. Konishi, and H. Suzuki, Genova preprint GEF-Th-7/1994, hep-th/9407027; C. Arvanitis, H.F. Jones, and C.S. Parker, Imperial College London preprint 1995 (hep-th/9502386) did not explain the exponentially fast convergence at strong couplings first found in Ref. [7]. This was explained in Ref. [8] and rigorously in

R. Guida, K. Konishi, and H. Suzuki, University of Genoa preprint, 1995 (hep-th/9505084).

[6] W. Janke and H. Kleinert, Phys. Lett. **A199**, 287 (1995),

[7] W. Janke and H. Kleinert, *Variational perturbation expansion for strong-coupling coefficients of the anharmonic oscillator*, Berlin preprint 1994 (quant-ph/9502019).

[8] H. Kleinert and W. Janke, *Convergence behavior of variational perturbation expansion — A method for locating Bender-Wu singularities*, Berlin preprint 1995 (quant-phys/9506160).

[9] H. Kleinert, *Variational Interpolation between Weak- and Strong-Coupling Expansions*, Berlin preprint 1995 (quant-phys/9507007).

[10] H. Kleinert, *Gauge Fields in Condensed Matter*,
Vol. I Superflow and Vortex Lines, pp. 1–744,
Vol. II Stresses and Defects, pp. 744-1443,
World Scientific, Singapore, 1989.

[11] J. Zinn-Justin, *Quantum Field Theory and Critical Phenomena*, Clarendon Press, Oxford, 1990.

[12] J. Zinn-Justin, J. Math. Phys. **22**, 511 (1981). This paper gives a correction factor to the imaginary part $[1 - (96/72)(3|g|) - (13259/10368)(3|g|)^2 - (8956043/2239488)(3|g|)^3 - 17.80162255(3|g|)^4 - 98.64510840(3|g|)^5 - 643.7460486(3|g|)^6 - \ldots ]$.

[13] H. Kleinert, Phys. Lett. **B300**, 261 (1993);
R. Karrlein and H. Kleinert, Phys. Lett. **A187**, 133 (1994).

[14] H. Kleinert, Phys. Lett. A **190**, 131 (1994).

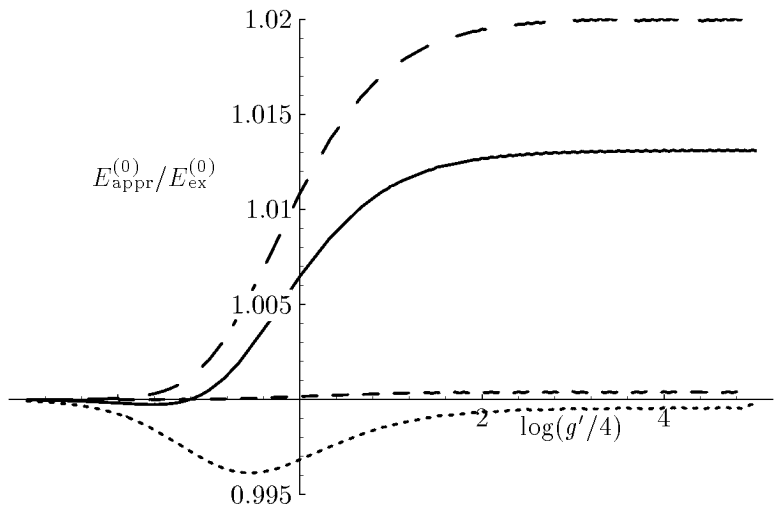

FIG. 1. Plot of the ratio of the resummed energy $E_1^{\text{appr}} = \bar{W}_1'(g)$ (solid curve) with respect to the exact energy as a function of the dimensionless coupling constant $g' = g/\omega^3$ for the anharmonic oscillator. The dashed curve shows the old approximation $W_1(g)$. The short-dashed curve indicates the approximation $W_3(g)$. The dotted curve indicates the approximation derived in Ref. [9] making use of the known strong-coupling behavior (which is not done here).

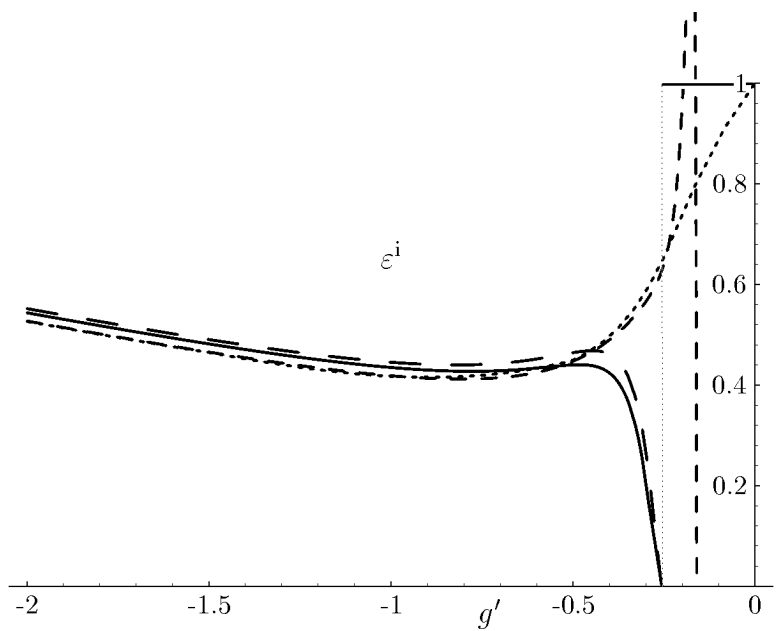

FIG. 2. The reduced imaginary part Im $\bar{W}_1'(g)$ (solid curve) of the ground-state energy in comparison with the exact one (dotted). The curve vanishes for $g \in (-0.26459, 0)$, where it is replaced by the semiclassical imaginary part. The dashed curve shows the approximation Im $W_1(g)$. The short-dashed curve indicates the approximation Im $W_3(g)$.

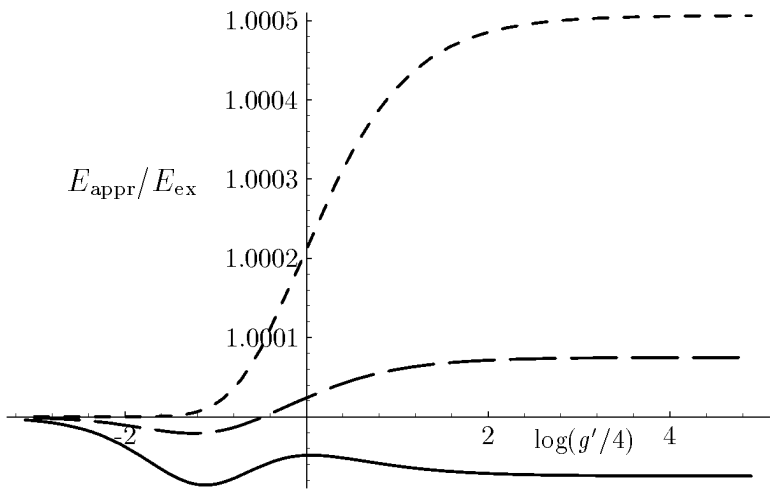

FIG. 3. The ratio of the approximation $E_2^{\mathrm{appr}} = \bar{W}_3'(g)$ with respect to the exact ground-state energy $E_{\mathrm{ex}}(g)$ of the anharmonic oscillator (solid curve). The short-dashed curve shows the old approximation $W_3(g)$, the long-dashed curve the approximation obtained for $g_3 = 0.16$ rather than the proper value $g_3 = 0.166$.

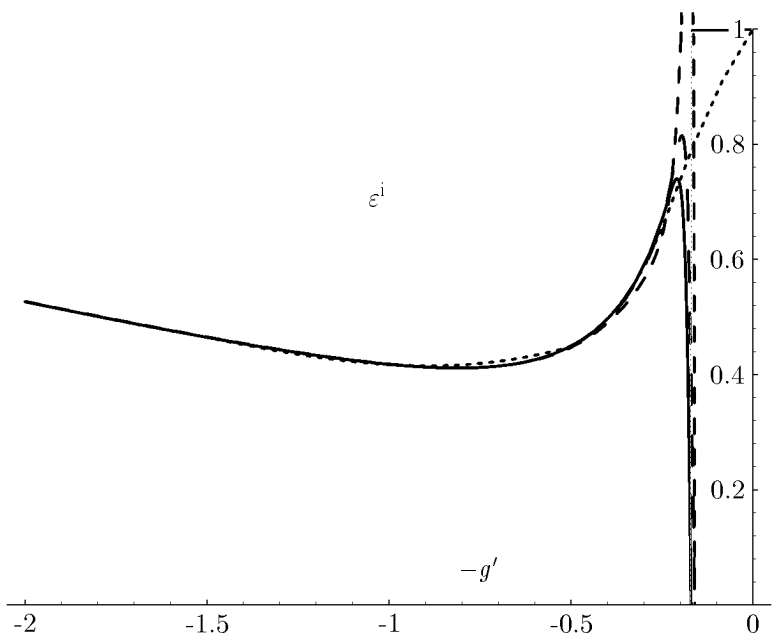

FIG. 4. The imaginary part of the approximation $\bar{W}_3'(g)$ (solid curve) vanishing for $g \in (-0.166, 0)$ where it is replaced by the semiclassical imaginary part. The dotted curve is the exact imaginary part. The dashed curve is obtained for $g_3 = 0.16$. The short-dashed curve shows the old approximation $W_3(g)$.

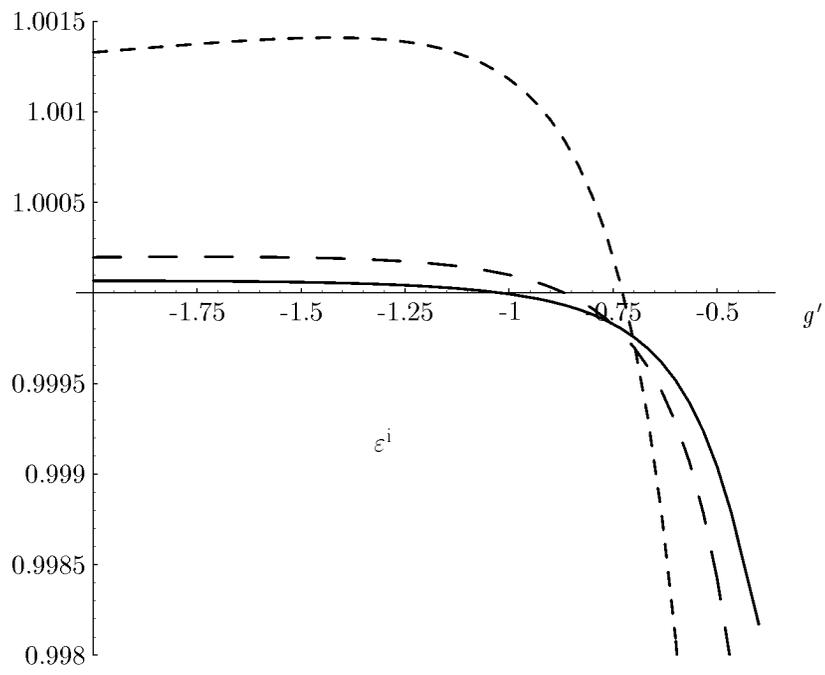

FIG. 5. Plot of the ratio of Im $W'_3(g)$ with respect to the exact Im $E_{\text{ex}}$ of the anharmonic oscillator as a function of $g' = g/\omega^3$ (solid curve). The dashed curve shows the old approximation $W_3(g)$, the short-dashed curve is obtained for $g_3 = 0.16$ rather than the proper value $g_3 = 0.166$.